# Effective Physical Processes and Active Information in Quantum Computing

Ignazio Licata
ISEM, Inst. For Scientific Methodology, PA, Italy
Ignazio.licata@ejtp.info

**Abstract**

The recent debate on hypercomputation has raised new questions both on the computational abilities of quantum systems and the Church-Turing Thesis role in Physics

We propose here the idea of "effective physical process" as the essentially physical notion of computation. By using the Bohm and Hiley active information concept we analyze the differences between the standard form (quantum gates) and the non-standard one (adiabatic and morphogenetic) of Quantum Computing, and we point out how its Super-Turing potentialities derive from an incomputable information source in accordance with Bell's constraints. On condition that we give up the formal concept of "universality", the possibility to realize quantum oracles is reachable. In this way computation is led back to the logic of physical world.

## 1. Introduction: "Purely Mechanical" (Turing, 1948)

One of the liveliest spheres in contemporary research is the study of the deep conceptual connection between Physics and Computation. Any physical system can be considered as an information processor continuously dialoguing with the external environment. The initial values are transformed into the final ones by the system's internal dynamics. The fundamental problem we deal with when working on such scenario is the role of Turing Computation in describing the informational activity of physical systems. The Church-Turing Thesis (CTT), in its strong form, states that any processing of syntactic information can be described by means of a suitable TM. To be more precise, any computation can be performed by a countably infinite collection of finite state automata by the name of UTM, Universal Turing Machine. From a mathematical viewpoint it is equivalent to define the computation concept by means of a TM and state that the actually countable functions are recursive functions. From the physical viewpoint, instead, the question is to give a precise meaning to the relation between the CTT and the world's description.

It is often maintained, quite imprecisely, that the CTT is a "statement on the physical world". Such claim is ambiguous and naïve at the same because physics does not deal with the "world in itself" but with the physical-mathematical models of it. So we have to inquire what Physics is at least compatible with a UTM. Some by now classical works (Church, 1957; Kolmogorov and Uspeskij, 1958; Kreisel, 1965; Gandy, 1980) have strengthen the idea that the behavior of a mechanical discrete system evolving according to local laws is recursive. These results can be extended both to the Markov Chains (Kreisel, 1970) and to a wide class of analog computers (Shannon, 1941; Pour-El, 1974; Lipshitz and Rubel, 1987). Such works have showed the relations between the Classical Computation Theory and determinism. In particular, it can be noted a strong analogy between a TM's globally unpredictable behaviors and deterministic chaos; in both cases the local rules do not imply a long-term predictable behavior, as a matter of fact.

The ordinary reasons providing a physical justification for the discrete and local features of TM are usually centered on some sort of atomism and the Relativity Principle. For example, it is possible to have recourse to the Bekenstein limit which fixes the amount of energetically distinguishable quantum states in a volume V (Bekenstein, 1981). A real physical system cannot read and manipulate more information than the one it can hold, so Turing assumption appears to be reasonable from a physical point of view. Furthermore the relativistic locality implies that the whole tape is not available in its entirety at each computational instant. Finally, the infinity of the tape can be seen as the consequence that no limitations are set to the possible implementations of the Kleene Theorem (for ex.: asynchronous and parallel computation, cellular automata and so on). It has to be noted that such argumentations make the recourse to the discrete structure of physical world sound more like the search for some sort of plausibility than a real reference to the dynamics of micro-physics.

So the TM remains a notion which was born within a classical and mechanistic conception of the physical world and the hilbertian axiomatics: "[TMs] can do anything that could be described as a 'rule of thumb' or 'purely mechanical', so that 'calculable by means of a [TM]' is the correct accurate rendering of such phrases" (Turing, 1948).

## 2. Effective Physical Processes as Computation

Roger Penrose (1989) proposed a physical version of the CTT: "*there exists a UTM whose repertoire includes any computation which can be performed by every physically possible object*". There are two different ways to interpret the Penrose thesis: we can take it as the possibility for a classical system to simulate a quantum one, which is a definitely controversial matter (Feynman, 2000), or to state that any function of a physical model is Turing-computable. Also the latter interpretation meets with many counter-statements even within the ambit of Classical Physics itself as well as in Quantum Cosmology (Pour-El and Richard, 1981; Geroch and Hartle, 1986; Costa and Doria, 1991; Scarpellini, 2002). There is a D. Deutsch's version (1997) which utilizes the "virtual reality" notion; but we can here express it in a much stronger way: *There exists a physical system whose evolution includes any physically possible system*. The benefit lies in bringing the whole debate back within Physics; in addition it is equivalent to stating a principle of self-similarity in the physical world such that a suitably "powerful" system is able to simulate any other one.

Clearly such kind of principle goes far beyond the original CTT, but it tends to *the redefining of the effective computability in terms of evolutive dynamics of the configurations of a physical system chosen as "universal" for reasons which are rooted in the experience of the physical world rather than for purely logical ones.* Thus the question of "universality" becomes strongly linked to the observer's choices and to the classes of physical environments into play. At this point it appears more natural adopting the idea of different CTTs for Physics and each one is valid under suitable conditions within definite ranges. For instance, the Fredkin and Toffoli "billiard-ball" machine (Fredkin and Toffoli, 1982) is a universal device able to display conservative logic and to account for mechanic processes; we can, otherwise, define relativistic computers whose computational resources make use of the typical topology of some space-time models; finally, under the hypothesis of "decoherent histories" (Griffiths, 2003; Gell-Mann and Hartle, 1993; Omnes, 1988, 1992,1994), it is possible to obtain the classical world structure as emergence. At this point, there are two key questions we have to answer to: A) is it possible to define an "all purpose" physical system able to act as a universal simulator of any other system? B) which are the "effective computability" peculiarities - meaning here "effective evolution"- of a physical system in relation to Turing-computability? The answer to the first question is presently controversial except we give it a banal reductionist meaning. Moreover, the real QFT capabilities to account for the mesoscopic emergent processes and the nature of noise are still matters under debate.

Such ideal system cannot surely be a classical one, because it is well known that a classical system can simulate a quantum one by undergoing, at least, exponential delay (Feynman, 2000), or by imposing drastic simplifications. What we have said about the essentially classical nature of TM seems to suggest its inaptness in simulating quantum systems. Thus, it is reasonable asking whether Quantum Computation (QC) displays Super-Turing abilities with respect to classical models, but the answer is paradoxical. In fact, if we adopt the idea of an "all purpose" scheme for a quantum net, then the benefit compared to classical computation is just in terms of efficiency, but if we let drop the universality condition for a QC, then we can get qualitatively different outcomes in respect to a TM. It has been recently showed that some performances of traditional QC can be de-quantized and simulated by a classical system (Calude, 2007, Aharonov et Al. 2007.). There are other research lines, we are here going to survey, which instead show hypercomputational abilities

In other words, there seem to exist two kind of quantum computation which we are going to analyze to better clarify the idea of "effective physical process" as physical computation.

### 3. Many Ways to Explore the Hilbert Hotel

The Quantum Computation history has developed along two main lines. On one hand the "auteur Hamiltonian", on the other one the quantum gates (Brown, 2000). The difference lies in the different relation with the search for universality from the logical viewpoint, and with decoherence from the physical viewpoint. Quantum networks are networks of standard computational steps provided by quantum gates which implement unitary operations on qbits. It is a high price to pay, because the temporal evolution of open quantum systems cannot fulfill the unitarity condition of operators. Thus, a perfect efficiency of quantum networks would imply a complete isolation of quantum gates from environment. Decoherence, instead, progressively erodes the superposition phase by deleting the off-diagonal elements in the density operator, so reducing the output to a statistical mixture.

For a Quantum Computer at thermic balance which processes a set of S qbits, the density matrix's off-diagonal elements exponentially decay according to (G.M. Palma et al., 1996):

$$\rho_{ij}(t) \approx \rho_{ij}(0) e^{-\gamma S t},$$

where $\gamma = 1/\tau_{dec}$ is a constant correlating the coupling to environment with the decoherence time. The crucial point is that the search for universality and the unitarity are deeply connected. The equivalent of a UTM for the QC is the existence of a universal "logic module" able to perform any computation by means of networks based on such element. It has been shown that if we limit ourselves to unitary operations on a finite number of qbits, such element will be a three-bit quantum gate or any non-trivial two-bit gate, like a Cphase gate (Deutsch, Barenco, Ekert, 1995; Serafini et al., 2006). In so doing we get a Quantum Universal Turing Machine (QUTM) which exploits the superposition principle and the entanglement among qbits as an element of non-classical parallelism within a "circuital" version still strongly based upon traditional models.

According to Bernstein and Vazirani (Bernstein and Vazirani, 1997) such QUTM can reduce classically exponential problems to quantum polynomial ones, and the standard quantum computation benefit would especially consist in efficiency. Besides decoherence and dissipation problems, this limiting outcome can be caused by the fact that the data acquisition is based on a process R (reduction) which operates as a selection among the superposition states, so providing outcomes in terms of classical bits.

There exists a different, and less constraining, approach to quantum computation, which makes use of the "controlled" evolution of the whole quantum system to get the sought answers. The model here is no more the universal, digital or quantum-digital-based one, but it is centered on the idea of analog quantum computer. There are two significant research lines to study the QC potentialities beyond Turing limit.

The first one is the Adiabatic Quantum Computation (AQC) (Farhi et al., 2000; Calude et al., 2000, 2002; Kieu, 2004; Zagoskin et al., 2007), centered on the evolution of a system described by a complex Hamiltonian where the problem evolving from a ground state to another is suitably codified, so as to give an answer by systematically exploring the Hilbert space. The adiabatic behavior of a quantum system (Messiah, 1999) warrants the possibility to approximate a non-perturbative situation, to obtain quite stable and not degenerated energy levels with excitement finite probabilities via Landau-Zener tunneling, as well as a "univocal" path from the initial state to the sought one in the space of the eigenstates occupied by virtue of the problem constraints in a finite time.

A satisfiability requirement for a n-bit system is generally a formula of the kind:

$$C_1 \wedge C_2 \wedge ... \wedge C_m$$

where all the $C_i$ are logical clauses. An AQC algorithm specifies an initial state within a set of n-qbits and a Hamiltonian $H(t)$ which describes its temporal evolution like this:

$$H(t) = H_{C1}(t) + H_{c2}(t) + .... + H_{cm}(t).$$

Each $H_{ci}$ is linked to the clause $C_i$, and for each $i$ between $0$ and $T$ the state $H_{ci}(t)$ codifies the satisfiability conditions for the correspondent clause and $H(T)$ the intersection ones for all the clauses.

The general AQC scheme is of the kind:

$$H(\lambda(t)) = H_f + \lambda(t) Z H_b,$$

where $H$ varies between the initial ground state $H_b$ and the final one $H_f$ with a Schrödinger-like evolution, $\lambda$ is the adiabaticity parameter and $Z \gg 1$. For a wide class of Hamiltonians it is possible to show that the escape probability of a state during the adiabatic evolution is given by a power law of the kind $|\dot\lambda|^\gamma$, with $\gamma = 1/2$ for all intermediate levels and $\gamma \leq 1/3$ for the edges.

An exemplar application of such method has been given by Tien Kieu (for ref. see: Kieu, 2006) in solving the Hilbert's tenth problem which is notoriously incomputable by traditional methods (Matiyasevich, 1993). Let's shortly recall the matter. It deals with finding a general algorithm to decide if a given Diophantine equation admits positive integer solution or not. Tien Kieu developed a quantum algorithm able to search within the domain of not negative integers for the absolute minimum of the square of a polynomial which is codified in an adiabatically evolving Hamiltonian. It corresponds to carry out a quantum oracle for the Hilbert's tenth problem by using the correspondence between not negative integers and occupation numbers in a Fock space. Given a Diophantine equation of the kind $D(x_1, x_2, ..., x_k) = 0$ we built a Hamiltonian of the kind $H_f = \left(D(a_1^+ a_1, ..., a_k^+ a_k)\right)^2$ whose observables in the ground-state will provide the sought answer in probabilistic terms.

There are several significant considerations to do about Kieu experience. We have firstly to notice that in this case the AQC replaces the classical concept of demonstration and performs computation as effective evolution of a physical system. As a matter of fact, the value of the procedure is "universal" for all Diophantine equations, but it is evident that universality is here connected to the type of adopted physics, and it is meant differently than under classical computation, *because the "algorithm" is the system itself.* Never during the process, either during elaboration or the final reading, the procedure's quantum nature is "forced", so obtaining the outcomes as probability distributions. In particular, the possibility to find a certain self-status of a final-time Hamiltonian within a dimensionally infinite space suggests that the system makes use of QM peculiar resources, such as tunnelling and interference. For instance, it is by means of AQC that Oh and Kim (Oh and Kim, 2007) have studied the GHZ entangled states for the phase transitions of a one-dimensional XY model and two-dimensional Ising model.

Another particularly interesting research line is that of quantum associative nets, which utilizes the quantum system abilities in image reconstruction and object recognition, two classical AI problems (Perus and Dey, 2000; Perus, 2000; Perus and Bischof, 2003; Perus, Bischof, Loo, 2004). It is a method which can ideally find its roots in the Davisson-Germer experiment and makes use of the ability to finely map the structures producing interference. It is an original quantum-wave implementation of Hopfield holographic nets whose efficiency is not comparable to the capabilities of the classical Hopfield nets. Also in this case we notice that the passage of a classical computational scheme within a different physics radically modifies its potentialities, because new processes, such as non-locality, come into play, while analogous proposals (Schutzhold, 2001) based upon quantum-implemented logic gates show much more limited, but - in some senses – more "universal" associative capacities. What we have seen for AQC can also be found in Quantum Morphogenic Computing (QMC) – as the abovementioned research is increasingly called - , i.e. a singular difference among the computational abilities depending on the way how QM is implemented. On one hand, the systems aspiring to universality by means of logical gates offer greater efficiency than the classical ones; on the other hand we find "dedicated"- and thus not-universal - quantum systems which heighten the QM potentialities, so confirming the idea of a "transversal" position of some quantum computation with respect to classical models.

Whether it is possible to reduce both AQC and QMC to the scheme of quantum gates is still an open problem. There are two fundamental problems. Firstly it has to be noticed that the response of the quantum gate is always classical, just because of the restrictions which unitarity and reversibility impose. Moreover, in following the Lloyd method (Lloyd, 1996) to connect AQC and QUTM, it is necessary to discretize the adiabatic computation time in order to express the adiabatic algorithm as a product of few q-bits. The essential point here does not only lie in the future technical implementations of q-gates, as it is erroneously believed, but in leading the intrinsically quantum features of adiabatic evolution – *which utilizes all the QM resources*! - back to the classical reading of outcomes.

## 4. Classical and Relativistic Hypercomputation

There are two kind of physical systems we take into consideration in this paper. The real physical systems - the actually realizable ones - such as the AQC, and the ideal physical systems, i.e. those compatible with some theoretical schemes, but not directly susceptible of implementation (*Gedankenexperiment*), whose function is to display the deep logic of the problem under consideration.

It can be useful here to shortly recall some classical and relativistic systems which showed beyond-Turing abilities. For instance, it has been demonstrated that - by using an opportune set of initial conditions - the systems of collinear or coplanar punctiform particles, subjected to classical gravity and elastic collisions (Mather and McGhee, 1975; Saari, 1977; Gerver, 1984; Gerver, 2003), can behave asymptotically so allowing a "billiard-balls"-like computation and being able to provide Super-Turing computational performances. A detailed argumentation would imply to take into consideration the theorems of uniqueness for global hyperbolicity in a Newton space-time. Here we will limit ourselves to point out that the core of the reasoning lies in demonstrating the existence of an infinite number of collisions – and in extracting an infinite quantity of energy from the gravitational potential as well – *in finite time*. If we associate a computational step to each collision it is possible to obtain a solution for the halting problem. The model's interest is purely theoretical because, quite patently, its exceptional characteristics for computation derive from ideal limit cases of Newtonian Physics.

The relativistic computers (Pitowsky, 1990; Eearman and Norton, 1993; Hogarth, 2004; Etesi and Nèmeti, 2002) are another instructive case; they utilize particular space-time topologies. The best-known case is the Malament-Hogarth space-time, where it is possible to make two computers working on different temporal scales and then to put them in communication each other. In fact, if there exists an infinite proper length worldline $\lambda$ which wholly belongs to the past of a certain event $p$, then an infinite computation along $\lambda$ can be used as oracle for the computation taking place on the worldline of $p$ in finite time. It is well-known that a TM getting infinite time at its disposal can solve the halting problem (Grunbaum, 1969; Hamkins and Lewis, 2000). Let's note that in this case the TM's capabilities do not derive from structural modifications, but from the possibility to compute in a sort of "present infinite" with respect to the $p$ point.

Although such kind of space-time is connected to physical "extreme" situations such as rotating charged black holes and cosmology - Kerr-Newman, Reiner-Nordström and anti-DeSitter solutions - we draw a particular interesting lesson for QC. The capacity to do infinite computation in finite time appears as a purely theoretical notion or to be linked to an exotic extension of some physical models, as a matter of fact it is a quite natural condition to define the essential characteristics of quantum computation.

### 5. The Role of Active Information in QC

The analysis of the QC different typologies leads to a paradoxical-tasting situation. In fact, the standard QC provides limited performances which fall under a traditional form of parallelism, while the AQC method and the similar ones have showed beyond-Turing performances. So it is necessary to ask whether a general characterization of the different QC forms is possible.

It has to be noticed that generally a finite-dimensional Hilbert space is used both in the logic quantum gates and in the AQC applied to the Hilbert's tenth problem. In the first case it corresponds to the necessity to take into consideration observables with finite spectra, such as Spin Systems or other kinds of quantum lattices. In the second case, Tien Kieu has referred to the connection between dimensionally finite Hilbert space and the logical nature of the problem which belongs to the class of the finitely refutable problems.

Such argumentations are justified by the adopted experimental set in the former case, and by the nature of the problem in the latter, but it is clear that the QC salient features do not depend on the Hilbert space finiteness, but they derive from the QM fundamental principles themselves. What QC does is to effectively and in natural way realize the classical hypercomputation conditions we saw in the paragraph 4, i.e. *exploring "many-worlds" in finite time by means of non-local performances!* The question can be understood if we attentively examine the quantum information dynamics in relation to the state preparation, the environment and the measurement procedures.

It can be useful here to refer to the active quantum information notion, used by Bohm and Hiley and the Birbeck College group (Bohm and Hiley, 2005; Hiley, Callaghan and Maroney, 2000; Hiley and Maroney, 1999, 2000; Hiley, 2002). Bohm approach to QM is formally equivalent to the standard one and it can provide a very refined epistemological scenario for the non-locality understanding.

As everybody knows, the non-local features can be described by the quantum potential:

$$Q(r,t) = -\frac{\hbar^2}{2m} \frac{\nabla^2 R(r,t)}{R(r,t)},$$

which derives from decomposing the Schrödinger Equation into real part and imaginary part, by using the polar expression $\Psi(r,t) = R(r,t)\exp[iS(r,t)/\hbar]$ for the wave function.

The quantum potential (QP) contains *in nuce* the essential features of QM, individuates an infinite set of phase paths for a quantum object and is responsible for entanglement. In particular, the QP has a contextual nature, i.e. it brings a global information on the process and its environment. The active information is defined as contextual constraint on the phase paths by the quantum potential. So it is the internal quantum information of the system, inaccessible to us, and unidentifiable with the Shannon information. Let's notice that such interpretation is absolutely general and can be naturally applied to the Feynman path integrals.

The quantum potential can be considered as an active information source linked to a quantum background (Implicate Order), which acts in the space-time where the measurements are made (Explicate Order). Implicate and Explicate order are connected by unfolding and enfolding processes which are defined through the Green's function and opportune parameters. By means of such construction Bohm and Hiley provided an operative version of the Bohr complementarity and derived the Heisenberg equations by means of the algebra of holomovements.

In Hiley words: *Shannon information will appear only when we consider a source that could be prepared in one of a number of orthogonal wave functions, each of which could be transferred separately. Here we have sufficient complexity to enable us to discuss channel capacities in terms of qbits in usual way.*

In other words we can obtain classical information only in the operations of state preparation and measurement, so making a selection of unfolded information from the Implicate Order, while the system's spontaneous activity directly operates with active, internal to the system and enfolded information in the Implicate Order.

By means of active information we meant to point out that the real Quantum Computing power can only be grasped within what lies at the bottom of QM, i.e. the Quantum Field Theory. For example, the working (in particular the outputs) of any quantum device is nothing but the outcome of a sort of phase transition and yet of not-balance. It is linked to the fact that such transition takes place in the presence of an external field interacting with the system in a not banal way. It means that the outputs bring inside, partly in implicit way, the external field' structure. Within such context, computation can't be controlled in traditional sense, but it becomes "emergent". Nevertheless, on occasion, under special circumstances, it is possible to exert peculiar control forms by acting on the environment-related macroscopic constraints and by codifying the problem on quantum states, it is just what Tien Kieu did.

We can so make the first physical distinction between the two forms of examined quantum computation by saying that the unitarity and reversibility constraints cause the quantum gate-like systems to work with unfolded information, and their computational power is thus limited to the outputs of the superposition states, while other methodologies let the system evolve according to its active enfolded information, so turning not only non-locality, but also decoherence, dissipation and the algorithmical uncompressibility of the probabilistic outcome into a resource.

## 6. In Search of a New Computational Paradigm for QC

Now it remains to be considered the general question about the computational abilities of a quantum system, especially as for the comparison with the classical theory of TMs.

The fidelity to the traditional computation model makes the QC of logical gates apt to define just a class of QP problems, i.e. solvable in polynomial time through a quantum algorithm. The problem of the AQC collocation appears to be more complex as well as, in a more general way, the one of the QC possibilities to compute beyond the Turing limit.

A global criticism of hypercomputation has been raised by Cotogno (Cotogno, 2003); it is related to the impossibility for any computational system to produce that "self-description" form which is the calculation of its own characteristic function. The analysis carried out by Ord and Kieu (Ord and Kieu, 2005; Kieu, 2004) has showed there is no contradiction in admitting that the diagonal functions cannot be computed by any function belonging to the same class on which they diagonalize, but they can be computed by a function belonging to a broader class. A different way to put the question - we can here only hint at it – is to consider a Turing oracle as a logical open system (Licata, 2007). But there is more than this. Ord and Kieu list a series of properties which a machine must possess to calculate its own characteristic function and they demonstrate that such properties are incompatible with each other. In other words the close relation between characteristic function and halting problem is only valid for recursive systems, but it does not define in itself a limitation to hypercomputation because this one could be the fruit of the system's peculiar resources which do not fall under the formal scheme of recursive systems. Once again the problem lies in taking into consideration the physics which supports the computational activity in relation to a specific problem and its codification. In particular Kieu has underlined the importance of the probabilistic reading of the outcomes, which thing makes QC fall outside the Cantor's diagonal argumentation.

We can understand the question by referring again to Bohm and Hiley's active information notion. In fact, if we consider any quantum phase path as an enfolded computational process, it becomes apparent that it is impossible to take such process into account within the classical computation theory which only "records" objective information in local and not-contextual way.

Ziegler analysis (Ziegler, 2005) points out how an infinite set of TMs working in parallel can solve the Hilbert's tenth problem and consequently the halting problem. If so, the QAC could formally fall under the traditional Turing computation scheme, in particular the class of semi-decidable problems.

Actually, such kind of equivalence fails to seize the substantial differences between the physical process and its formal description. In fact, the active information of a quantum system is defined by an uncountable finite number of phase paths and has a strictly non-local nature. Even if a future Quantum Gravity Theory succeeds – how the Bekenstein limit seems to show – in identifying a limit on Planck scale, it must not furnish grounds for identifying the activity of a single "space-temporal cell" with the TM one. Moreover it has to be noticed that semi-decidability fully finds its physical significance in the fact that the system explores the active information space in search of equilibrium situations which display themselves by means of probability distributions.

In the QC general case the non-Turing features directly derive *from the QM structure itself*. In fact, active information is intrinsically not-computable; if it were so, it would mean to violate the Bell Theorem on the impossibility of a QM with local hidden variables.

The QC hypercomputational potentialities thus derive from the "unbounded" active information role in acting as "oracular source" in particular experimental configurations.

## 7. Conclusion. Quantum Oracles and Universality

The active information notion provides a useful conceptual tool to distinguish the QC standard forms (logical gates on q-bits) from the non-standard ones, and allows of easily fitting the non-Turing features which emerge in quantum systems without any contradiction to the theory formal structure. It is the QM structure itself which gives evidence for the not-computable informational content in physical world. In this way computation by effective physical processes show to be broader than the Turing one, and only a future post-QFT will be able to clarify the definitive limits of such kind of computation.

The hypercomputational performances in QC are thus connected to the realization of oracles which utilize the active information of quantum systems. It depends on the kind of problem under consideration and its codification in particular experimental devices.

So it seems that the price to pay in order to make use of the QC hypercomputational potentialities is to give up its logical-formal universality, so leading back the computation notion to the physical world's logic.

Scarpellini, B. (2002),*Two Indecidable Problems of Analysis*, in *Mind and Machines,*13,1,pags.49-77

Serafini, A.S. Bose, A.S., Mancini, S.(2006), *Distributed Quantum Computation via Optical Fibres*, in *Phys. Rev. Lett.* 96, 010503

Shannon,C.E., (1941), *Mathematical Theory of the Differential Analyser,* in *Jour. of Math. And Phys. Of the MIT,*20,pags. 337-354

Schutzhold, R. (2001), *Pattern Recognition on a Quantum Computer*, e.print in arXiv:quant-ph/0208063

Turing, A.M. (1948), *Intelligent Machinery - National Physical Laboratory Report*, in B. Meltzer B., D. Michie, D. (eds) 1969, *Machine Intelligence* 5, Edinburgh University Press

Zagoskin, A.M.,Savel'ev S., Nori, F. (2007), *Modeling an Adiabatic Quantum Computer via an Exact Map to a Gas of Particles*, in *Phys. Rev. Lett.* 98, 120503

Ziegler, M. (2005), *Computational Power of Infinite Quantum Parallelism*, in *Int. Journ. of Theor. Phys.,*44, 11, pags.2059-2071
11